\documentstyle[sprocl,axodraw,psfig]{article}

\newcommand{\gsim}{ \mathop{}_{\textstyle \sim}^{\textstyle >} }

\def\Journal#1#2#3#4{{#1} {\bf #2}, #3 (#4)}

\def\NPB{{\em Nucl. Phys.} B}
\def\PLB{{\em Phys. Lett.}  B}
\def\PRL{\em Phys. Rev. Lett.}
\def\PRD{{\em Phys. Rev.} D}

\def\be{\begin{equation}}
\def\ee{\end{equation}}
\def\bea{\begin{eqnarray}}
\def\eea{\end{eqnarray}}

\begin{document}

\title{ Neutrino Oscillation and Charged Lepton-Flavor Violation
       \\ in the Supersymmetric Standard Models  }

\author{J.~Hisano$^{\rm a}$ and D.~Nomura$^{\rm a,b}$}
\address{{\rm a)} Theory Group, KEK, Tsukuba, \\
Ibaraki 305-0801, Japan\\
{\rm b)} Department of Physics, University of Tokyo, \\
Tokyo 113-0033, Japan\\ 
E-mail: junji.hisano@kek.jp, dnomura@ccthmail.kek.jp}

\maketitle

\abstracts{ The neutrino experiment results suggest that the neutrinos
have finite masses and the lepton-flavor symmetries are violating in
nature.  In the supersymmetric models, the charged lepton-flavor
violating processes, such as $\mu^+\rightarrow e^+ \gamma$ and
$\tau^\pm \rightarrow \mu^\pm \gamma$, may have the branching ratios
accessible to the future experiments, depending on origins of the
neutrino masses and the SUSY breaking. In this paper we discuss the
branching ratios in the supergravity scenario using the current solar
and atmospheric neutrino experimental data.}

\section{Introduction}

Result of the atmospheric neutrino experiment by the superKamiokande
detector indicates that the neutrinos have finite masses and the
lepton-flavor symmetry of muon is violating in nature \cite{atm}. This
is the first signature of the physics beyond the standard model (SM),
and this discovery will be confirmed by further experiments, such as
the long base-line experiments. Also, the solar neutrino experimental
data suggest that the lepton-flavor symmetry of electron is
violating \cite{solar}.

Processes, such as $\mu^+\rightarrow e^+ \gamma$ and $\tau^\pm
\rightarrow \mu^\pm \gamma$, are also lepton-flavor violating (LFV)
processes. Unfortunately, the event rates are too small to be observed
in near future experiments even if the neutrino masses are introduced
into the standard model. The event rates are suppressed by the fourth
order of the ratio of the tiny neutrino mass to the $W$ boson mass due
to the GIM suppression. However, if the standard model is
supersymmetrized, the processes may be accessible in near future
experiment and we may study the origin of the neutrino masses.

The supersymmetric standard model (SUSY SM) is a solution of the
naturalness problem, and is one of the most promising model beyond the
standard model. In this model, introduction of the SUSY breaking terms
allows the lepton-flavor symmetries to be violating in the slepton
masses \cite{EN}. Then, the orders of magnitude of the event rates for
the LFV processes depend on the origin of the SUSY breaking in the
SUSY SM and physics beyond the SUSY SM. One of the successful
candidates for the origin of the SUSY breaking is the minimal
supergravity, where the SUSY breaking scalar masses are generated
universally in the flavor space at the tree level.  In this scenario, the
source of the LFV processes comes from the LFV radiative correction to
the SUSY breaking masses for the sleptons by the LFV interaction in
physics beyond the SUSY SM \cite{HKR}. Then, we have a chance to study
the origin of the neutrino masses through the LFV processes in the
supersymmetric models.

The see-saw mechanism \cite{seesaw} is the simplest model to generate
the tiny neutrino masses. In this mechanism the Yukawa interactions
are lepton-flavor violating due to introduction of the right-handed
neutrinos, similar to the quark sector. Then, in the minimal
supergravity scenario, if the lepton-flavor violation in the Yukawa
coupling constants is strong enough, the radiative correction
generates sizable LFV masses for the sleptons \cite{BM}\cite{HMTYY}.
Moreover, the large mixing angles observed on the solar and
atmospheric neutrino observations may enhance the event rates for the
LFV processes \cite{HMTY}\cite{HNY}\cite{HN}.

In this article, we study the charged lepton-flavor violating
processes, $\mu^+\rightarrow e^+ \gamma$ and $\tau^\pm \rightarrow
\mu^\pm \gamma$, using the current neutrino experimental data, in the
SUSY SM with the right-handed neutrinos.\footnote{
After sleptons are discovered in the future 
large colliders, the slepton oscillation will be a powerful tool to 
study the lepton-flavor violation  \cite{sleptonosc}.
} 
We assume the minimal supergravity scenario.  The large mixing angle
in the atmospheric neutrino result may enhance $\tau^\pm \rightarrow
\mu^\pm \gamma$, and the large mixing angles in the MSW and the vacuum
oscillation solutions may lead to a large event rate of
$\mu^+\rightarrow e^+ \gamma$.

This article is organized as follows.  In the next section we review
the radiative generation of the LFV masses for slepton in the SUSY SM
with the right-handed neutrinos. In Section 3 we show the branching
rate for $\tau^\pm \rightarrow \mu^\pm \gamma$, using the atmospheric
neutrino result. In Section 4 discuss the branching rate for $\mu^+
\rightarrow e^+ \gamma$, using the solar neutrino result. The other
processes are also discussed. Section 5 is devoted to Conclusion.

\section{The SUSY SM with the right-handed neutrinos}

We review the radiative generation of the LFV masses for sleptons in
the SUSY SM with the right-handed neutrinos. We adopt the minimal
supergravity scenario as the origin of the SUSY breaking in the SUSY
SM. The superpotential of the lepton sector in the SUSY SM with
right-handed neutrinos is given as
\begin{eqnarray}
W &=& f_{\nu_{ij}} H_2 \overline{N}_i L_j 
                 +f_{e_{ij}} H_1 \overline{E}_i L_j 
                 + \frac12 M_{\nu_i\nu_j} \overline{N}_i \overline{N}_j,
\label{yukawacoupling}
\end{eqnarray}
where $L$ is a chiral superfield for the left-handed lepton, and
$\overline{N}$ and $\overline{E}$ are for the right-handed neutrino
and the charged lepton.  $H_1$ and $H_2$ are for the Higgs doublets in
the SUSY SM. Here, $i$ and $j$ are generation indices.  After
redefinition of the fields, the Yukawa coupling constants and the
Majorana masses can be taken as
\begin{eqnarray}
f_{\nu_{ij}}&=&f_{\nu_i} V_{D ij},\nonumber\\
f_{e_{ij}}&=&f_{e_i}\delta_{ij}, \nonumber\\
M_{\nu_i\nu_j} &=& U_{ik}^{\ast} M_{\nu_{k}} U^\dagger_{kj},
\end{eqnarray}
where $V_D$ and $U$ are unitary matrices. 
In this model the mass matrix for the left-handed neutrinos
$(m_{\nu})$ becomes
\begin{eqnarray}
(m_{\nu})_{ij} &=& 
V_{D ik}^\top (\overline{m}_{\nu})_{kl} V_{D lj},
\label{mbar}
\end{eqnarray}
where 
\begin{eqnarray}
(\overline{m}_{\nu})_{ij} &=& 
m_{{\nu_i}D} \left[M^{-1}\right]_{ij} m_{{\nu_j}D} 
\nonumber\\
&\equiv& V_{Mik}^\top m_{\nu_k}  V_{Mkj}.
\end{eqnarray}
Here, $m_{{\nu_i}D}=f_{\nu_i}v\sin\beta/\sqrt{2}$
and $V_M$ is a unitary matrix.\footnote{ 
$\langle h_1 \rangle=(v\cos\beta/\sqrt{2},0)^\top$ and
$\langle h_2 \rangle=(0,v\sin\beta/\sqrt{2})^\top$ with $v\simeq 246$GeV.}
The observed mixing angles on the atmospheric and solar neutrino
experiments are $(V_M V_D)_{\tau\mu}$ and $(V_M V_D)_{\mu e}$, respectively,
if they come from the oscillations of $\nu_\mu-\nu_\tau$ and
$\nu_e-\nu_\mu$.

The SUSY breaking terms for the lepton sector in the SUSY SM
with the right-handed neutrinos are in general given as
\begin{eqnarray}
-{\cal L}_{\rm \ SUSY \ breaking}&=& \phantom{+}
 (m_{\tilde L}^2)_{ij} \tilde{l}_{Li}^{\dagger} \tilde{l}_{Lj} 
+(m_{\tilde e}^2)_{ij} \tilde{e}_{Ri}^\ast \tilde{e}_{Rj}  
+(m_{\tilde \nu}^2)_{ij} \tilde{\nu}_{Ri}^\ast \tilde{\nu}_{Rj}  
\nonumber \\
& &
 + (A_\nu^{ij} h_2 \tilde{\nu}^\ast_{Ri} \tilde{l}_{Lj}
   +A_e^{ij} h_1 \tilde{e}^\ast_{Ri} \tilde{l}_{Lj} 
   +\frac12 B_{\nu}^{ij}  \tilde{\nu}^\ast_{Ri} \tilde{\nu}^\ast_{Rj} 
    + h.c.),
\label{MSSMsoft}
\end{eqnarray}
where $\tilde{l}_L$, $\tilde{e}_R$, and $\tilde{\nu}_R$ represent the
left-handed slepton, and the right-handed charged slepton, and the
right-handed neutrino. Also, $h_1$ and $h_2$ are the doublet Higgs
bosons. In the minimal supergravity scenario the SUSY breaking masses
for sleptons, squarks, and the Higgs bosons are universal at the
gravitational scale ($M_{\rm grav} \sim 10^{18}$GeV), and the SUSY
breaking parameters associated with the supersymmetric Yukawa
couplings or masses ($A$ or $B$ parameters) are proportional to the Yukawa
coupling constants or masses.  Then, the SUSY breaking parameters in
Eq.~(\ref{MSSMsoft}) are given as
\begin{eqnarray}
&(m_{\tilde L}^2)_{ij}=(m_{\tilde e}^2)_{ij}= (m_{\tilde \nu}^2)_{ij}
=\delta_{ij} m_0^2 ,& \nonumber\\ 
&A_\nu^{ij} = f_{\nu_{ij}}a_0,~~~A_e^{ij} = f_{e_{ij}} a_0,& \nonumber\\ 
&B_{\nu}^{ij} = M_{\nu_i \nu_j} b_0,&
\end{eqnarray}
at the tree level. 

In order to know the values of the SUSY breaking parameters at the low
energy, we have to include the radiative corrections to them. While we
evaluate them by solving the RGE's, we discuss only the qualitative
behavior of the solution using the logarithmic approximation here.
The SUSY breaking masses of squarks, sleptons, and the Higgs bosons at
the low energy become heavier by gauge interactions at one-loop
level, and the corrections are flavor-independent. On the other hand,
Yukawa interactions reduce the diagonal SUSY breaking mass squareds
and the radiative corrections are flavor-dependent. Then, if the
Yukawa coupling is lepton-flavor violating, the radiative correction
to the SUSY breaking parameters is also lepton-flavor violating.  The
LFV off-diagonal components for $(m_{\tilde L}^2)$, $(m_{\tilde
e}^2)$, and $A_e$ in the SUSY SM with the right-handed neutrinos
are given at the low energy as
\begin{eqnarray}
(m_{\tilde L}^2)_{ij}&\simeq&
-\frac1{8\pi^2} 
(3m_0^2+a_0^2) 
V_{D ki}^\ast V_{D lj} f_{\nu_k} f_{\nu_l} U_{km}^\ast U_{lm}
\log \frac{M_{\rm grav}}{M_{\nu_m}},
\nonumber\\
(m^2_{\tilde e})_{ij} &\simeq&   0,
\nonumber\\
A_e^{ij} &\simeq& 
-\frac{3}{8\pi^2} a_0
f_{e_i} V_{D ki}^\ast V_{D lj} f_{\nu_k} f_{\nu_l} U_{km}^\ast U_{lm}
\log \frac{M_{\rm grav}}{M_{\nu_m}},
\label{LFVinMSSMRN}
\end{eqnarray}
where $i\ne j$. In these equations, the off-diagonal components of
$(m_{\tilde L}^2)$ and $A_e$ are generated radiatively while those of
$(m_{\tilde e}^2)$ are not. This is because the right-handed leptons
have only one kind of the Yukawa interaction $f_e$ and we can always
take a basis where $f_e$ is diagonal. This is a characteristic of the
SUSY SM with the right-handed neutrinos.
\footnote{
In the minimal SU(5) SUSY GUT, the right-handed sleptons receive the
LFV masses through the Yukawa interaction of colored Higgs, but not the
left-handed ones \cite{BH} \cite{HMTYG}. In the SO(10) SUSY GUT and
the non-minimal SU(5) SUSY GUT both sleptons may have the LFV masses
\cite{others}.
} 

The magnitudes of the off-diagonal components of $(m_{\tilde L}^2)$
and $A_e$ are sensitive to $f_{\nu_i}$ and $V_D$, while not to $U$. This
is because the off-diagonal components of $U$ are small when the
hierarchy among the right-handed neutrino masses is large, and then we
will take $U={\bf 1}$ in the following discussion for simplicity. In
the following sections we will evaluate the values of $f_{\nu_i}$ and
$V_D$ from the neutrino oscillation data.

\section{The atmospheric neutrino result and  $\tau^\pm\to\mu^\pm\gamma$}

\begin{figure}
\begin{center}
\begin{picture}(120,80)
\ArrowArcn(67.5,12.5)(37.5,180,135)
\ArrowArc(67.5,12.5)(37.5,90,135)
\ArrowArcn(67.5,12.5)(37.5,90,45)
\ArrowArc(67.5,12.5)(37.5,0,45)
\Vertex(67.5,50){3}
\Text(67.5,62.5)[]{$v_2$}
\Vertex(94,39){3}
\Vertex(41,39){3}

\Text(15,39)[]{$\tilde{H}^-(\tilde{H}^0)$}
\Text(120,39)[]{$\tilde{W}^-(\tilde{W}^0)$}

\ArrowLine(30,12.5)(15,12.5)
\DashArrowLine(30,12.5)(67.5,12.5){3}             \Vertex(67.5,12.5){3}
\DashArrowLine(67.5,12.5)(105,12.5){3}        
\ArrowLine(105,12.5)(120,12.5)

\Text(22.5,2.5)[]{$\tau_R$}
\Text(48.5,2.5)[]{$\tilde{\nu}_\tau(\tilde{\tau}_L)$}
\Text(86,2.5)[]{$\tilde{\nu}_\mu(\tilde{\mu}_L)$}
\Text(112.5,2.5)[]{$\mu_L$}

\Text(67.5,25)[]{$(m^2_{\tilde{L}})_{\tau\mu}$}

\Photon(86,55)(97.5,67.5){2}{5}
\Text(101.5,72.5)[]{$\gamma$}

\end{picture} 
\end{center}
\caption{The Feynman diagram which gives a dominant contribution to
$\tau^+ \to \mu^+ \gamma$ when $\tan\beta\gsim 1$ and the off-diagonal
elements of the right-handed slepton mass matrix are negligible, as in
the MSSM with the right-handed neutrinos.  $\tilde{\tau}_{L}$ and
$\tilde{\mu}_{L}$ are the left-handed stau and smuon, respectively,
and $\tilde{\nu}_{\tau}$ and $\tilde{\nu}_{\mu}$ the tau sneutrino and
the mu sneutrino.  $\tilde{H}$ is Higgsino, $\tilde{W}$ wino.  The
arrows represent the chiralities.}
\end{figure}
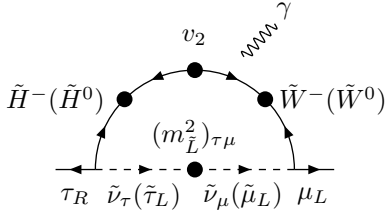

\begin{figure}
\begin{center}
\centerline{\epsfxsize=4.0cm \epsfbox{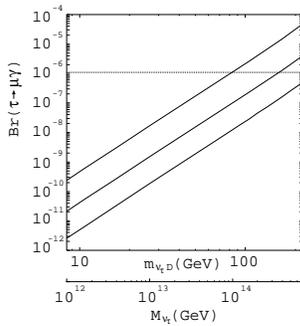}}
\vspace*{-5mm}
\caption{Dependence of the branching ratio of $\tau\to\mu\gamma$ on
the Dirac neutrino mass for the tau neutrino $m_{\nu_{\tau} D}$ ( the
right-handed tau neutrino mass $M_{\nu_\tau}$). The dotted line is the
current experimental bound.
Here, $m_{\nu_\tau}=0.07$eV, $\sin2\theta_D=1$. Also, we take
$m_{\tilde{e}_L} =170$GeV and the wino mass 130GeV.  The other gaugino
masses are determined by the GUT relation for the gaugino masses for
simplicity. Also, we impose the radiative breaking condition of the
SU(2)$_L\times$U(1)$_Y$ gauge symmetries with $\tan\beta=3,10,30$ and
the Higgsino mass parameter positive.  Here also the larger
$\tan\beta$ corresponds to the upper line.}
\end{center}
\end{figure}

In this section we discuss the branching ratio of $\tau^\pm\to\mu^\pm\gamma$
using the atmospheric neutrino result. From the zenith-angle dependence
of $\nu_e$ and $\nu_{\mu}$ fluxes measured by the superKamiokande
it is natural that the atmospheric neutrino anomaly comes from the 
neutrino oscillation between $\nu_\mu$ and $\nu_\tau$, and 
the neutrino mass-squared difference and mixing angle are
expected as 
\begin{eqnarray}
&\Delta m^2_{\nu_{\mu} \nu_\tau} \simeq 10^{-(2-3)} {\rm eV^2}, &
        \nonumber\\
&\sin^2 2\theta_{\nu_{\mu} \nu_\tau}  \gsim 0.8.&
\end{eqnarray}
Assuming that the neutrino masses is hierarchical as $m_{\nu_{\tau}}\gg
m_{\nu_{\mu}} \gg m_{\nu_{e}}$, the tau neutrino mass is given as
\begin{eqnarray}
m_{\nu_\tau} \simeq (3 \times 10^{-2} - 1 \times 10^{-1}){\rm eV},
\label{taumass}
\end{eqnarray}
and if the tau neutrino Yukawa coupling constant $f_{\nu_\tau}$ is as
large as that of the top quark, the right-handed tau neutrino
$M_{\nu_\tau}$ is about $10^{14-15}$GeV.

In order to evaluate the event rate for $\tau^\pm\to\mu^\pm\gamma$, we
have to know the value of $V_{D\tau\mu}$, which is not necessary the same
as the $\sin \theta_{\nu_{\mu} \nu_\tau}$. However, it is expected
that it is also of the order of one as explained bellow.

Let us consider only the tau and the mu neutrino masses for
simplicity. In this case we parameterize two unitary matrices $V_D$
and $V_M$ as
\begin{eqnarray}
V_D= 
\left(
\begin{array}{cc} 
\cos\theta_D& \sin\theta_D \\
-\sin\theta_D& \cos\theta_D
\end{array}
\right),&&
V_M= 
\left(
\begin{array}{cc} 
\cos\theta_M&  \sin\theta_M \\
-\sin\theta_M& \cos\theta_M
\end{array}
\right).
\end{eqnarray}
The observed large angle $\theta_{\nu_{\mu} \nu_\tau}$ is a sum of
$\theta_D$ and $\theta_M$. However, in order to derive $\theta_M\sim
\pi/4$ we need a fine-tune among the independent Yukawa coupling
constants and the mass parameters.  The neutrino mass matrix
$(\overline{m}_{\nu})$ for the second and the third generations
(Eq.~(\ref{mbar})) is given  as
\begin{eqnarray}
(\overline{m}_{\nu}) &\propto& 
\left(
\begin{array}{cc}
\frac{m_{\nu_{\mu} D}^2}{M_{\nu_{\mu} \nu_{\mu}}}&
-\frac{m_{\nu_{\mu} D}m_{\nu_{\tau} D}}{M_{\nu_{\mu}\nu_{\tau}}} \frac{M_{\nu_{\mu}\nu_{\tau}}^2}{M_{\nu_{\mu}\nu_{\mu}}M_{\nu_{\tau}\nu_{\tau}}}\\
-\frac{m_{\nu_{\mu} D}m_{\nu_{\tau} D}}{M_{\nu_{\mu}\nu_{\tau}}} \frac{M_{\nu_{\mu}\nu_{\tau}}^2}{M_{\nu_{\mu}\nu_{\mu}}M_{\nu_{\tau}\nu_{\tau}}}&
\frac{m_{\nu_{\tau} D}^2}{M_{\nu_{\tau}\nu_{\tau}}} 
\end{array}
\right).
\end{eqnarray}
If the following relations are valid, 
\begin{equation}
\frac{m_{\nu_{\tau} D}^2}{M_{\nu_{\tau}\nu_{\tau}}} 
\simeq \frac{m_{\nu_{\mu} D}^2}{M_{\nu_{\mu}\nu_{\mu}}}
\simeq \frac{m_{\nu_{\mu} D} m_{\nu_{\tau} D}}{M_{\nu_{\mu}\nu_{\tau}}},
\label{relation}
\end{equation}
$m_{\nu_\tau} \gg m_{\nu_\mu}$ and $\theta_M\simeq \pi/4$ can be
derived.  However, the relation among the independent coupling
constants and masses is not natural without some mechanism or
symmetry.  Also, if $m_{\nu_{\tau} D}\gg m_{\nu_{\mu} D}$ similar to
the quark sector, the mixing angle $\theta_M$ tends to be  suppressed as
\begin{equation}
\tan 2 \theta_M \simeq
2 \left(\frac{m_{\nu_{\mu} D}}{m_{\nu_{\tau} D}} \right)
  \left(\frac{M_{\nu_{\mu}\nu_{\tau}}}{M_{\nu_{\mu}\nu_{\mu}}}     \right).
\end{equation}
Therefore, in the following discussion we assume that the large mixing
angle between $\nu_\tau$ and $\nu_\mu$ comes from $\theta_D$ and that
$V_M$ is a unit matrix.

Large $V_{D \tau\mu}$ leads to the non-vanishing $(m_{\tilde
L}^2)_{\tau\mu}$ and $A_e^{\tau\mu}$, which result in a finite
$\tau^\pm\to\mu^\pm\gamma$ event rate via diagrams involving them.
They are given as
\begin{eqnarray}
(m_{\tilde L}^2)_{\tau \mu}&\simeq&
\frac1{16\pi^2} (3m_0^2+a_0^2) 
  \sin 2 \theta_D f_{\nu_{\tau}}^2  \log \frac{M_{\rm grav}}{M_{\nu_{\tau}}},
\nonumber\\
A_{e}^{\tau \mu}&\simeq&
\frac3{16\pi^2} a_0
  \sin 2 \theta_D f_{\tau} f_{\nu_{\tau}}^2  \log \frac{M_{\rm grav}}{M_{\nu_{\tau}}} .
\end{eqnarray}
As will be shown, if $f_{\nu_{\tau}}$ is of the order of one, the branching
ratio of $\tau\to \mu\gamma$ may reach the present experimental bound.

Let us evaluate the branching ratios of $\tau\to\mu\gamma$. The
amplitude of the $e_i^+\to e_j^+\gamma$ ($i>j$) takes a form
\begin{eqnarray}
T=e \epsilon^{\alpha*}(q) \bar{v}_{i} (p) 
i \sigma_{\alpha \beta} q^\beta (A^{(ij)}_L P_L + A^{(ij)}_R P_R)
v_{j}(p-q),
\label{Penguin}
\end{eqnarray}
where $p$ and $q$ are momenta of $e_i$ and photon, and the event rate
is given by
\begin{eqnarray}
\Gamma(e_i \to e_j\gamma)
= \frac{e^2}{16 \pi} m_{e_i}^3 (|A^{(ij)}_L|^2+|A^{(ij)}_R|^2).
\label{eventrate}
\end{eqnarray}
Here, we neglect the mass of $e_j$.  The amplitude is not invariant
for the SU(2)$_L$ and U(1)$_Y$ symmetries and the chiral symmetries of
leptons. Then the coefficients $A^{(ij)}_L$ and $A^{(ij)}_R$ are
proportional to the charged lepton masses. Since the mismatch between
the left-handed slepton and the charged lepton mass eigenstates is
induced in the SUSY SM with the right-handed neutrinos, $A^{(ij)}_L$
is much larger than $A^{(ij)}_R$. Also, when $\tan\beta(\equiv
v_2/v_1)$ is large, the contribution to $A^{(ij)}_L$ proportional to
$f_{e_i} v_2 (= - \sqrt{2} m_{e_i} \tan\beta)$ becomes dominant.
Then, the dominant contribution to $\tau^\pm \to\mu^\pm\gamma$ is from
the diagram of Fig.~(1) for $\tan\beta\gsim 1$.

In Fig.~(2) we show the branching ratio of $\tau^\pm\to\mu^\pm\gamma$
as a function of the Dirac neutrino mass for tau neutrino
$m_{\nu_{\tau} D}$ (the right-handed tau neutrino mass
$M_{\nu_\tau}$). Here, $m_{\nu_\tau}=0.07$eV, $\sin2\theta_D=1$. Also,
we take $m_{\tilde{e}_L} =170$GeV and the wino mass 130GeV.  The other
gaugino masses are determined by the GUT relation for the gaugino
masses for simplicity. Also, we impose the radiative breaking
condition of the SU(2)$_L\times$U(1)$_Y$ gauge symmetries with
$\tan\beta=3,10,30$ and the Higgsino mass parameter positive. The
branching ratio is proportional to $m_{\nu_{\tau} D}^4$
($M_{\nu_\tau}^2$).  The current experimental bound is $Br\le
1.1\times 10^{-6}$ \cite{cleo}, and some region is excluded by it. If
$10^{-8}$ can be reached in the future experiments, such as B
factories, we can probe $m_{\nu_{\tau} D}> 20(80)$GeV for
$\tan\beta=30(3)$. Then, if the Dirac tau neutrino mass is as large as
the top quark mass, we may observe $\tau^\pm\to\mu^\pm\gamma$ there.

\section{The solar neutrino result and  $\mu^+\to e^+ \gamma$}

\begin{figure}
\begin{center}
\centerline{\epsfxsize=8.0cm \epsfbox{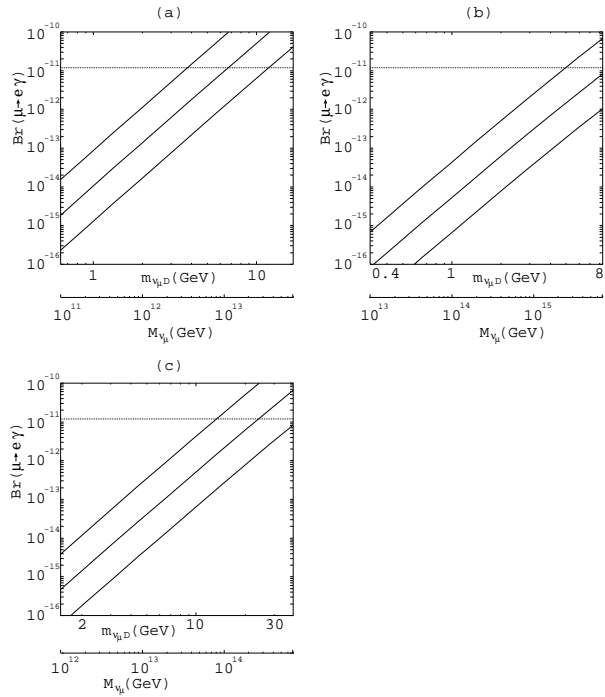}}
\caption{Dependence of the branching ratio of $\mu\to e\gamma$ on the
Dirac neutrino mass for the mu neutrino $m_{\nu_{\mu} D}$ (the
right-handed mu neutrino mass $M_{\nu_\mu}$). Here, a) is for the MSW
solution with the large angle, b) is the vacuum oscillation solution,
and c) is for the MSW solution with the small angle. $m_{\nu_\mu}$ and
$V_D$ are given in text.  The other parameters are the same as in
Fig.~(2). The dotted lines are the current experimental bound. }
\end{center}
\end{figure}

In this section we discuss the relation between the solar neutrino
result and $\mu^+\to e^+ \gamma$, assuming that the solar neutrino
deficit comes from the  $\nu_e-\nu_\mu$ oscillation. The relation is
more complicated compared with that between the atmospheric neutrino
result and $\tau^\pm\rightarrow\mu^\pm\gamma$.

There are three candidates for the solution of the solar neutrino deficit
 if it comes from neutrino oscillation. The MSW solution \cite{MSW}
 due to the matter effect in the sun gives the natural explanation,
 and the observation favors
\begin{eqnarray}
&\Delta m^2_{\nu_e \nu_Y} \simeq 10^{-(4-5)} {\rm eV^2} {~~~\mbox{ or}~~~} 10^{-7}
{\rm eV^2}, &
        \nonumber\\
&\sin^2 2\theta_{\nu_e \nu_Y} \gsim 0.5, &
\end{eqnarray}
or
\begin{eqnarray}
&\Delta m^2_{\nu_{e} \nu_Y} \simeq 10^{-5}
{\rm eV^2}, &
        \nonumber\\
&\sin^2 2\theta_{\nu_e \nu_Y} \simeq 10^{-(2-3)} .&
\end{eqnarray}
If the solar neutrino anomaly comes from so-called 'just so' solution
\cite{justso}, the neutrino oscillation in vacuum, the mass-squared
difference and mixing angle are expected as \cite{justso}
\begin{eqnarray}
&\Delta m^2_{\nu_{e} \nu_Y} \simeq 10^{-(10-11)}{\rm eV^2}, &
        \nonumber\\
&\sin^2 2\theta_{\nu_{e} \nu_Y}  \gsim 0.5 .&
\end{eqnarray}
Assuming that the neutrino masses hierarchical as
$m_{\nu_{\tau}}\gg m_{\nu_{\mu}} \gg m_{\nu_{e}}$, it is natural to
consider $\nu_Y = \nu_{\mu}$. If one of the large angle solutions
for the solar neutrino anomaly is true, the large mixing
$\theta_{\nu_{\mu}\nu_e}$ may imply the LFV large mixing for sleptons
between the first- and the second-generations.  Similar to the
atmospheric neutrino case, it is natural to consider that the large
mixing angle between $\nu_\mu$ and $\nu_e$ in the MSW solution or the 'just
so' solution for the solar neutrino anomaly comes from $V_D$.

The amplitude for $\mu^+\rightarrow e^+ \gamma$ is proportional to 
$(m_{\tilde L}^2)_{\mu e}$, and it has two contributions
in the SUSY SM with right-handed neutrinos as 
\begin{eqnarray}
(m_{\tilde L}^2)_{\mu e} &\simeq&
-\frac1{8\pi^2} (3m_0^2+a_0^2) \times
\nonumber\\
&&
 \left(
  V_{D \tau \mu}^\ast V_{D \tau e} f_{\nu_{\tau}}^2  \log \frac{M_{\rm grav}}{M_{\nu_{\tau}}} 
+ V_{D \mu \mu}^\ast V_{D \mu e} f_{\nu_{\mu}}^2  \log \frac{M_{\rm grav}}{M_{\nu_{\mu}}} 
\right).
\label{m12}
\end{eqnarray}
Here, we assume $f_{\nu_{\tau}}\gg f_{\nu_{\mu}} \gg f_{\nu_{e}}$, and
the term proportional to $f_{\nu_{e}}^2$ is neglected.  Unfortunately,
we do not have information about $V_{D \tau e}$ and we can not
evaluate the first term in Eq.~(\ref{m12}). On the other hand, we can
evaluate the second term if $V_{D \mu e}$ can be determined from the
solar neutrino result.  Then, in the following, we evaluate the event
rate for $\mu^+\rightarrow e^+ \gamma$ assuming $V_{D\tau e}=0$.
Notice that though this gives the conservative value for the event
rate, there are also possibilities where the event rate is larger or
smaller due to the finite $V_{D \tau e}$.

Let us evaluate $\mu^+\to e^+ \gamma$. The forms of the amplitude and
the event rate are the same as those of $\tau^\pm\to\mu^\pm\gamma$
(Eqs.~(\ref{Penguin},\ref{eventrate})). As mentioned above, if the solar
neutrino anomaly comes from the MSW effect or the vacuum oscillation
with the large angle, $V_{D \mu e}$ is expected to be large. This may
lead to large $(m_{\tilde L}^2)_{\mu e}$.  In Fig.~(3-a), under the
condition that
\begin{equation}
V_{D} =\left( \begin{array}{ccc}  
 \frac{1}{\sqrt{2}}   &         \frac12      &   \frac12            \\
-\frac{1}{\sqrt{2}}   &         \frac12      &   \frac12            \\
            0         & -\frac{1}{\sqrt{2}}  &  \frac{1}{\sqrt{2}} 
\end{array} \right), \label{eq:BiMax} 
\end{equation}
we show the branching ratio of $\mu^+\to e^+\gamma$ as a function of
$m_{\nu_\mu D}$ ($M_{\nu_\mu}$). We take $m_{\nu_\mu}=4.0\times
10^{-3}$eV, which is consistent with the MSW solution. The other input
parameters are taken to be the same as in Fig.~(2). The branching
ratio is promotional to $m_{\nu_\mu D}^4$ ($M_{\nu_\mu}^2$). For
$\tan\beta=30(3)$, the branching ratio reaches the experimental bound
(${\rm Br}(\mu\to e\gamma) < 1.2\times10^{-11}$ \cite{MEGA}) when
$m_{\nu_\mu D}\simeq4 (10)$GeV. A future experiment at PSI is expected
to reach 10$^{-14}$ \cite{kuno}.  This corresponds to $m_{\nu_\mu
D}\simeq 0.5 (2)$GeV. If we take $m_{\nu_\mu}=1.0\times 10^{-5}$eV
expected by the 'just so' solution (Fig.~(3-b)), the branching ratio
becomes slightly smaller for a fixed $m_{\nu_\mu D}$ since the log
factor in Eq.~(\ref{m12}) is smaller.

If the solar neutrino anomaly comes from the MSW solution with the
small mixing, we cannot distinguish whether the mixing comes from
$V_D$ or $V_M$ even if using argument of naturalness.  If it comes
from $V_D$, the branching ratio is smaller by about 1/100 compared
with that in the MSW solution with the large mixing, as shown in
Fig.~(3-c).  In Fig.~(3-c) we assume
that
\begin{equation}
V_{D} =\left( \begin{array}{ccc}  
  1    &   0.04  & 0.03            \\
- 0.04 &   0.79  & 0.59            \\
  0    & - 0.60  & 0.80
\end{array} \right) \label{eq:MSWSmallMixing} 
\end{equation}
and $m_{\nu_\mu}=2.2\times 10^{-3}$eV. Other input parameters are the same as
Fig.~(2).

Finally we consider the $\mu^+ \to e^+ e^- e^+$ process and the
$\mu$-$e$ conversion in nuclei.  For these processes the penguin type
diagrams tend to dominate over the others, so the behavior of the
decay rate is similar to that of $\mu^+ \to e^+ \gamma$. For the
$\mu^+ \to e^+ e^- e^+$ process the following approximate relation
holds,
\begin{eqnarray}
{\rm Br}(\mu \to 3e) &\simeq& \frac{\alpha}{8\pi} \frac83  
        \left(  \log \frac{m_\mu^2}{m_e^2} - \frac{11}{4}\right)
       {\rm Br}(\mu \to e \gamma) \\
 &\simeq& 7 \times 10^{-3} {\rm Br}(\mu \to e \gamma).
\end{eqnarray}
For the $\mu$-$e$ conversion rate $\Gamma(\mu \to e)$
a similar relation holds at the $\tan\beta\gsim 1$ region,
\begin{equation}
\Gamma (\mu \to e) \simeq 16\alpha^4 Z_{\rm eff}^4 Z |F(q^2)|^2
                           {\rm Br}(\mu \to e \gamma) .
\end{equation}
Here $Z$ is the proton number in the nucleus, and $Z_{\rm eff}$ is the
effective charge, $F(q^2)$ the nuclear form factor at the momentum
transfer $q$. The $\mu$-$e$ conversion rate normalized by the muon
capture rate in Ti nucleus, $R(\mu^- \to e^- ;{}^{48}_{22}{\rm Ti})$,
is approximately
\begin{equation}
R(\mu^- \to e^- ;{}^{48}_{22}{\rm Ti}) 
\simeq 6 \times 10^{-3} {\rm Br}(\mu \to e \gamma) .
\end{equation}
The MECO collaboration proves that they have a technology to reach
$R(\mu^- \to e^- ; {}^{48}_{22}{\rm Ti})<10^{-16}$
\cite{meco}. Furthermore, now there are active discussions of the high
intensity muon source, and we may reach to a level of $10^{-18}$ if
the muon storage is constructed and $10^{(19-20)}$ muons per a
year are produced\cite{HIMS}. This is comparable to ${\rm Br}(\mu\rightarrow
e\gamma) \sim 10^{-16}$, and we can probe the region $m_{\nu_\mu
D}\simeq 0.2 (0.5)$GeV ($M_{\nu_\mu} \simeq 10^{10}$ $(10^{11})$GeV)
in the MSW solution with the large angle.

\section{Conclusion}

In this article we discuss the charged lepton-flavor violating
processes, $\mu^+\rightarrow e^+ \gamma$ and $\tau^\pm \rightarrow
\mu^\pm \gamma$, using the current neutrino experimental data, in the
SUSY SM with the right-handed neutrinos.  While this model has many
unknown parameters, these processes may be accessible in near future
experiment. The LFV search will give new insights to the origin of the
neutrino masses.

\section*{References}

\end{document}